# Morphing for faster computations in transformation optics

**Ronald Aznavourian[1],* and Sébastien Guenneau[1]**

[1]*Aix-Marseille Université, CNRS, Ecole Centrale Marseille, Institut Fresnel, 13013 Marseille, FRANCE*
*ronald.aznavourian@fresnel.fr

**Abstract:** We propose to use morphing algorithms to deduce some approximate wave pictures of scattering by cylindrical invisibility cloaks of various shapes deduced from the exact computation (e.g. using a finite element method) of scattering by cloaks of two given shapes, say circular and elliptic ones, thereafter called the source and destination images. The error in $L^2$ norm between the exact and approximate solutions deduced via morphing from the source and destination images is typically less than 1 percent if control points are judiciously chosen. Our approach works equally well for rotators and concentrators, and also unveils some device which we call rotacon since it both rotates and concentrates electromagnetic fields. However, it breaks down for superscatterers (deduced from non-monotonic transforms): The error in $L^2$ norm is about 25 percent. We stress that our approach might greatly accelerate numerical studies of 2D and 3D cloaks.

## 1. Introduction

In the early 90's, there was a very fashionable special effect called "morphing" which consisted in transforming an image into another with a succession of intermediate images. This computer graphic technique [1], notably used in the heroic fantasy movie *Willow* and the musical video clip *Black or White* of Michael Jackson is fairly underrated in other contexts, such as photonics. However, the morphing approach is reminiscent of transformational optics (TO), whereby a coordinate change stretches some Cartesian grid (or more generally a Delauney mesh) in a way similar to what researchers do to control light trajectories in transformed (metamaterial) media [2,3,4]. In what follows, we would like to show that morphing can deepen our understanding of TO by adopting a different viewpoint.

Of all the sciences, there's one, the computer science, which is used on a daily basis by researchers in miscellaneous fields, including in photonics (but also biological sciences etc). Although it may seem to be at first glance a world apart from physical sciences-in the sense that it's a transversal science, a kind of tools factory used in different professions-we shall see it can be an invaluable help for people working on TO. Indeed, by combining physical and computer sciences, we are led to an interesting way of working with the morphing.

In section 2, we describe the morphing principle and how it works in practice through a first example in cloaking. In section 3, we describe the limit of applicability of morphing to TO in more details through a superscattering (space folding) example (reminiscent of cloaking via anomalous resonances [5,6]), as well as illustrate how morphing can be used as a guide towards design of novel TO metamaterials. In section 4, we draw some concluding remarks.

## 2. Morphing: Principle and first example

*2.1 Principle*

« Morphing » is the word used to say: « image transformation ». If at the beginning, the morphing was just a simple interpolation of colors, with the source's image colors which progressively fade away to let (also progressively) appear the destination's image colors, we shall see that nowadays the morphing is slightly more complicated. In fact, morphing is now based on a double interpolation, both on shapes and on colors, between two images. There are different ways to do this, and we have chosen to look into only the most wide-spread methods. To illustrate our discussion, we have used a free software called: "Sqirlz Morph" [7].

*2.2 Control points*

Whichever morphing technique we use, we need « control points ».

1. Definition: The « control points » are « points » manually placed by the user, in both (source and destination) images, to determinate the most important parts of these images, and more particularly, the parts in one image that should exactly be placed in the right position in the other image. We have to specify that at each control point in one image corresponds one control point in the other one, in order to establish a one-to-one map.

2. Importance: Control points are essential for the shape's transformations. Indeed, keeping in mind that morphing preserves the proportions, we can easily understand that if there are some parts of one image which are exactly placed at the right position in the other image, then there's a high probability that the whole transformation will be right. In fact, the more control points, the higher the probability to get it right.

3. Constraints: To be useful, control points should be neither aligned nor too close from each other. The number of control points is also important. Indeed, too few control points will produce a superposition of the source and destination images, instead of a real transformation of one into the other. On the other hand, too many control points might produce antagonistic transformations and spoil the result. So, placing the control points requires some thinking, and cannot be easily automated [8]. This is in essence the price to pay to achieve faster results: Any human intervention means a more subjective result.

*2.3 Circular and elliptical cloaks as a first example*

We stress the importance of control points in Figures 1 & 2, where we consider a circular cloak of inner radius $a_1=0.2$m and outer radius $b_1=0.3$m, an incident transverse electric plane wave from the left at wavelength $\lambda=0.25$m, which we map on an elliptical cloak [9] of semi-axes $a_1=0.2$m, $b_1=0.4$m (inner elliptical boundary) and $a_2=0.4$m, $b_2=0.8$m (outer elliptical boundary). One can see that when we take more control points (judiciously located), the result of morphing compared against the finite element numerical result for an elliptical cloak of eccentricities $a_1=0.2$m, $b_1=0.3$m and $a_2=0.4$m, $b_2=0.6$m is improved.

We recall the geometric transform for elliptical cloaks [5]:

$$f:(r,\theta) \longrightarrow (r',\theta') = (\alpha r + \beta, \theta), \text{ where } \forall 0 < r < R_2 = \sqrt{a_2^2 \cos^2\theta + b_2^2 \sin^2\theta}$$

$$\alpha = (R_2 - \beta)/R_2, \text{ with } \beta = \sqrt{a_1^2 \cos^2\theta + b_1^2 \sin^2\theta},$$

which maps the area within the ellipsis of eccentricities $a_2$, $b_2$ onto an elliptical corona delimited by ellipses of eccentricities $a_1$, $b_1$ and $a_2$, $b_2$.

One can then deduce the transformed tensors of permittivity $\varepsilon'$ and permeability $\mu'$ in the transformed coordinates from the computation of the Jacobian matrix $J(r',\theta') = \partial(r,\theta) / \partial(r',\theta')$, and the coefficients of the metric tensor $T = J^T J / det(J)$ through the formula $\varepsilon' = \varepsilon T^{-1}$, $\mu' = \mu T^{-1}$, where $\varepsilon$, $\mu$ are the (possibly heterogeneous but isotropic) permittivity and permeability of the medium in the original coordinate system. It is well known that T is infinitely anisotropic at the cloak's inner boundary, and this leads to some numerical inaccuracies. We thus decide to set either homogeneous Dirichlet or Neumann data at the cloak's inner boundary (which corresponds to infinite conducting boundary in the transverse magnetic, or transverse electric polarization, respectively). We first compute with the commercial finite element package COMSOL multiphysics the scattering of circular and elliptical cloaks making infinite conducting circular and elliptical inclusions unseen by a transverse electric plane wave, see fig. 1. Note that in Fig. 1 we chose two extreme cases whereby the number of control points is either too small (a,b) or too large (c,d).

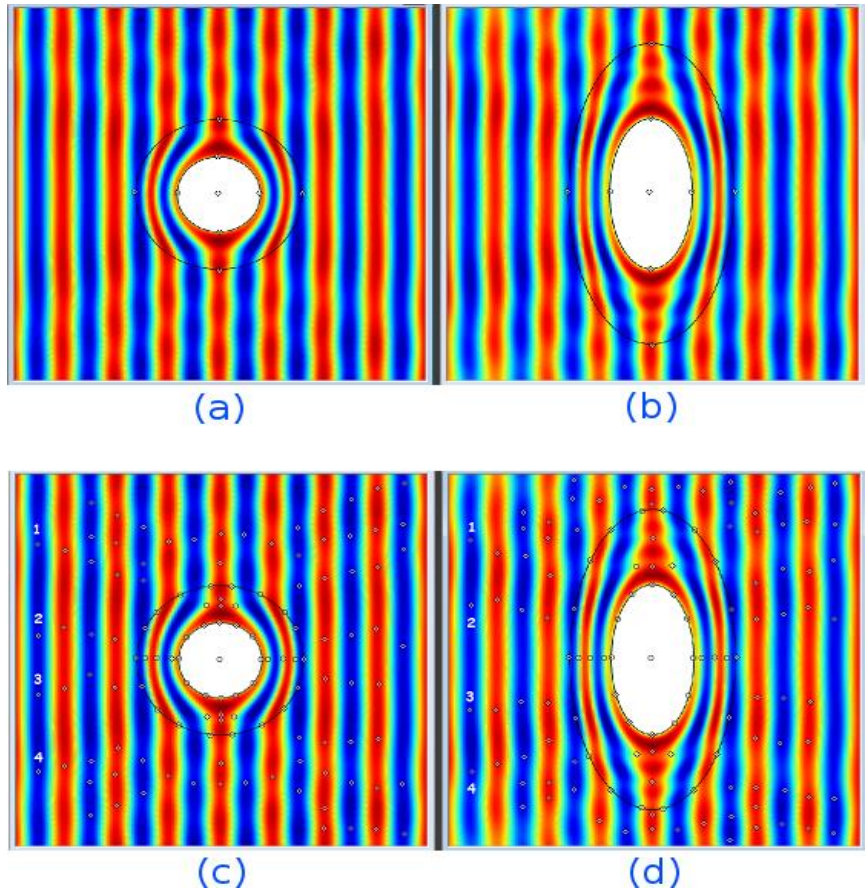

Fig. 1. Case of morphing from (a) to (b) and from (c) to (d) with too few control points in (a) and (b) and too many control points in (c) and (d), with also some control points aligned and close to other ones in (c) and (d). One can see that the control point numbers "1", "2", "3", "4" in (c) correspond to the same control points in (d), with however a stretch along the vertical axis.

As a result of morphing (50%) between the source image Fig.1(a,c) and destination image Fig.1(b,d), we show in Fig. 2(a) a rather crude approximation by the morphing from Fig. 1(a) to Fig. 1(b), and a better approximation in Fig. 2(b) by the morphing from Fig. 1(c) to Fig. 1(d). We show for comparison the result of a direct computation with Comsol Multiphysics in Fig. 2(c).

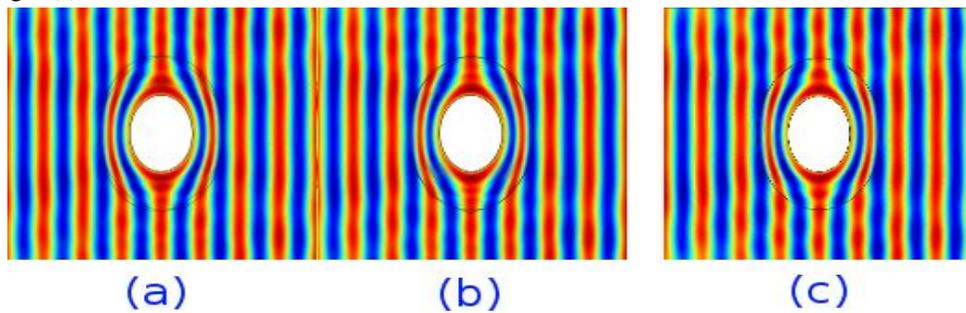

Fig. 2. Obtained intermediate result (image half-way from the final image) in the case of morphing with a few number of control points in (a), and with too many control points in (b). The result of a direct computation in (c) is quite close from the morphing result in (b).

In conclusion of this section, we may say that the secret of a good morphing work lies in the number and in the judicious positions of the control points.

*2.4 The two main methods*

We have seen the importance of the control points. It remains to create a meshing with these control points to process the transformation of the source image to the destination image. There are different methods to perform this operation, and we'll see now the two main ones.

1. The mesh warping: This method is based on a kind of rectangular meshing of the two images which mustn't be totally different, or which must be topologically equal. Then it works in two steps:

• First step, we create a temporary array of meshes which have the same ordinates as the source image and the same abscissors as the destination image. Then, for every ordinate we calculate the spline function between the source image's abscissors and the temporary array of mesh's abscissors.

• Second step, for every abscissors we calculate the spline function between the temporary array of mesh's ordinates and the destination image's ordinates.

In this way, with the spline's functions, we can determinate for each point in the source image the coordinates of this point in the destination image.

2. The Delaunay's triangulation:

This method is based on a triangular meshing of the two images, following the Delaunay's triangulation rules (see Fig.3 for an example of warp and triangular meshings). Then, every triangle in the source image is transformed into a triangle in the destination image. Note that even if there are no problems with vertices of triangles, a difficulty might arise from points located inside the triangles. To overcome such a problem, we use the Voronoï's diagram which allows us to determinate the coordinates of points inside the triangles from the source image to the destination image.

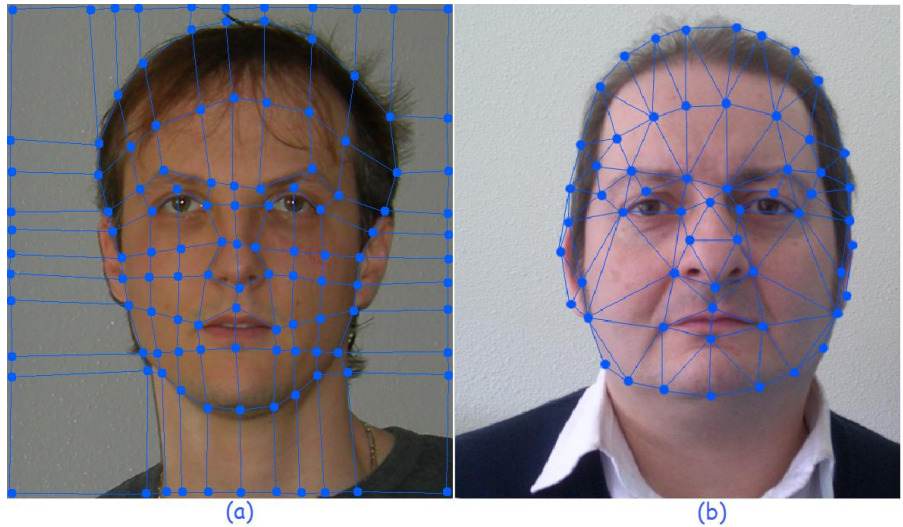

Fig. 3. Rectangular (warp) meshing example in (a) and triangular (Delaunay) meshing example in (b).

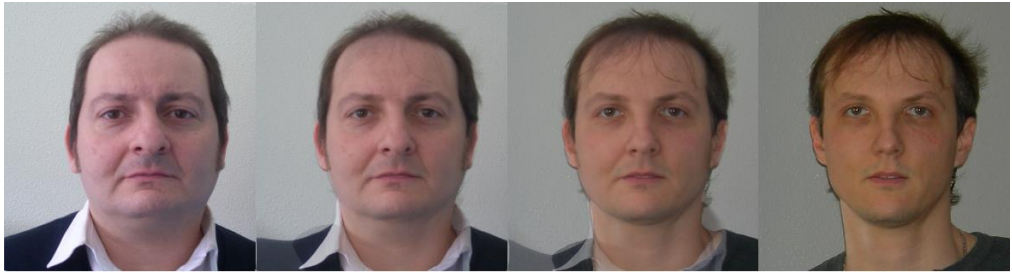

Fig. 4. Morphing example with a mapping from the leftmost into the rightmost photos, with intermediate images obtained with respectively 33% and 66% of morphing. One should note that both shapes and colors are interpolated.

*2.5 How to use the morphing tool*

First of all, you need two images of computed results which mustn't be totally different: one need consider images with some similarities i.e. two faces, two invisibility cloaks and so forth. The overall size and shape of images matters. For instance, if one would like to get morphing results between invisibility cloaks and human faces as is done in Fig. 2 and 4, the first thing to check is to equalize the size of source and destination images.

Then, you need of course a morphing software in order to process the morphing transformation. We opted for the freeware Sqirlz Morph [7].

With this software, one just has to place the control points in one image, making sure that each control point in the first image is correctly placed in the final image. Note that one can move each control point in one image independently from its corresponding control point in the other image. One just has to perform the morphing transformation by using this software, and pick up the intermediate morphing transformation image one wants in order to obtain the desired result.

**3. Strength and weakness of morphing as a tool for transformation optics**

*3.1 Application field*

Now that we know how morphing works, we can envisage many possible applications, and not only in an optical context. In fact, morphing can be applied to other wave phenomena provided we consider linear governing equations (what corresponds to linear transforms). In our case (an electromagnetic wave application), as we will see, morphing proves to be already an invaluable tool.

*3.2 Interest*

We have seen how morphing is working, and how one can use this tool. The main interest of this renewed approach to TO is to considerably reduce computational time.

But let's get this straight: we still need someone to compute the starting and destination images and even more importantly someone who can judiciously place control points therein, before using any morphing algorithm.

*3.3 Efficiency*

Until now, morphing has shown some interesting features, but the real question is to estimate the efficiency of this tool. By naked eyes, looking at the results in figure 2, we are tempted to claim it is indeed efficient, but this remains a qualitative feeling: We need a quantitative numerical answer. So we proceed in two stages, the first one consists in working with images in gradation of grey, with white for the highest value, and black for the lowest one, then we make the difference between the morphing result image and the fully computational result image, in such a way that the darker a pixel, the lesser the difference. On the contrary, the brighter a pixel, the larger the discrepancy between images at this pixel location. In the second stage, we apply a $L^2$ norm function adapted to the field size, on all the pixels of the difference image, so that we obtain a value between 0 and 1, with 0 when the compared images are exactly the same, and 1 when the compared images are totally different. Note that this comparison function is not linear and behaves as a square root function. With this method, we obtained the numerical value 0.05 for the comparison between the two images with the difference image (c) of fig.5.

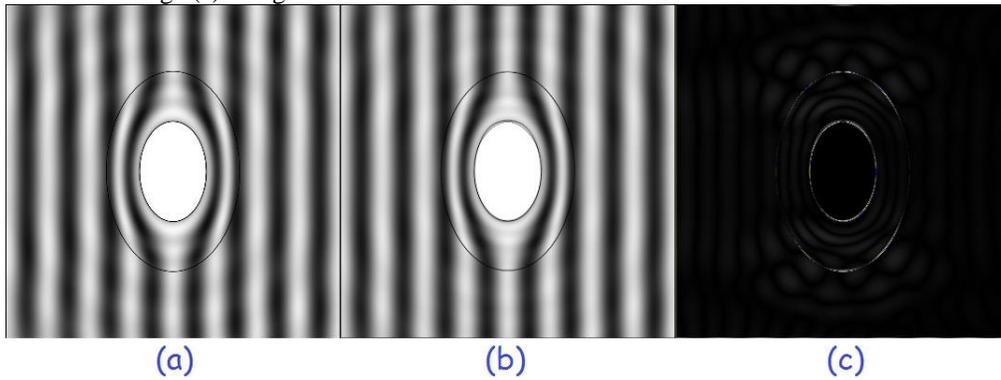

Fig. 5. Comparison by difference (c), in gradation of grey, between a calculated result (a) and a morphing result (b).

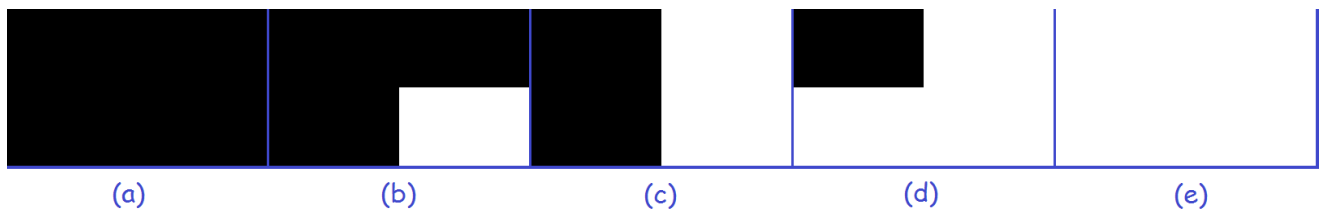

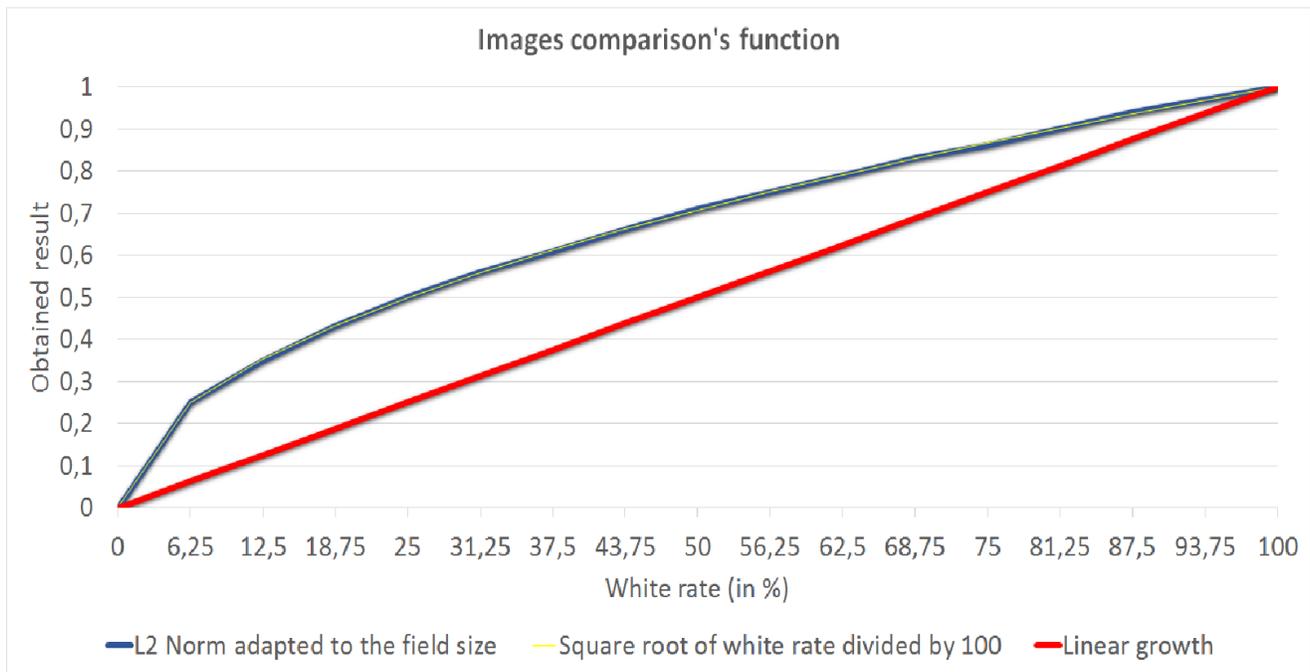

Fig. 6. Test of L² norm for image comparison's function (f) with black and white images with respectively 0% (a), 25% (b), 50% (c), 75% (d), and 100% (e) of white.

*3.4 Limits of applicability of morphing*

Even if the morphing's technique has proved its efficiency with the monotonous functions, and, in particular, with the linear functions, on the other hand, this technique is no longer efficient as soon as we apply it to a non monotonous function. To illustrate this statement, we choose to test the morphing's technique on superscatterer cloaks which are designed via non monotonous functions. These are deduced from space folding transforms [10]. More precisely, the transformed permittivity and permeability have the following values:

$\varepsilon' = \varepsilon (R_2)^4/r^4$, $\mu' = -\mu$ where $\varepsilon$ and $\mu$ are the permittivity and permeability of the medium surrounding the cloak of outer radius R2.

In our examples, we consider in Fig. 7(a) a circular superscatterer cloak of inner radius $R_1=0.25$m, with outer radius $R_2=0.4$m, which scatters like a perfect magnetic conducting (PMC) circular obstacle of radius $R_3=0.64$m, see Fig. 8(a), for a transverse electric plane wave incident from the top at wavelength $\lambda=0.14$m. We show in Fig. 7(b) a circular superscatterer cloak of inner radius $R_1=0.1$m, with outer radius $R_2=0.2$m, which scatters like a PMC obstacle of radius $R_3=0.4$m, see Fig. 8(b), with the same incident transverse electric plane wave as in Fig. 7(a) and Fig. 8(a). For the intermediate size cloak in Fig. 7(d), we have considered a circular cloak of inner radius $R_1=0.175$m, with outer radius $R_2=0.3$m and $R_3=0.5142857$m (that is with inner and outer radii averages of those in Fig. 7(a) and Fig. 7(b)), always with the same incident transverse electric plane wave. We then compare these fully numerical solutions with those reported in Fig. 7(c) (resp. Fig. 8(c)), which are obtained by 50% morphing between Fig. 7(a) (resp. Fig. 8(a)), the source image, and Fig. 7(b) (resp. Fig. 8(b)), the destination image.

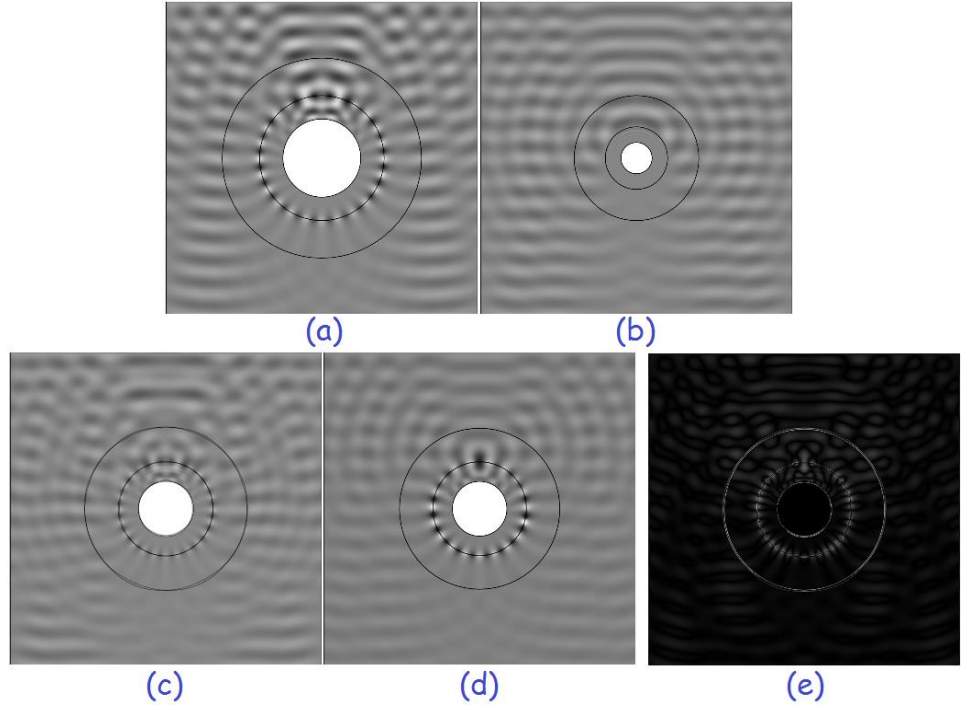

Fig. 7. Superscatterer cloaks with inner radii $R_1=0.25$m (a) and $R_1=0.1$m (b) and outer radii $R_2=0.4$m (a) and $R_2=0.2$m (b). The inner discs are filled with infinite conducting material. Obtained intermediate result (image half-way from the final image) in the case of morphing in (c), and a direct computation in (d). In (e) we show the difference image between (c) and (d).

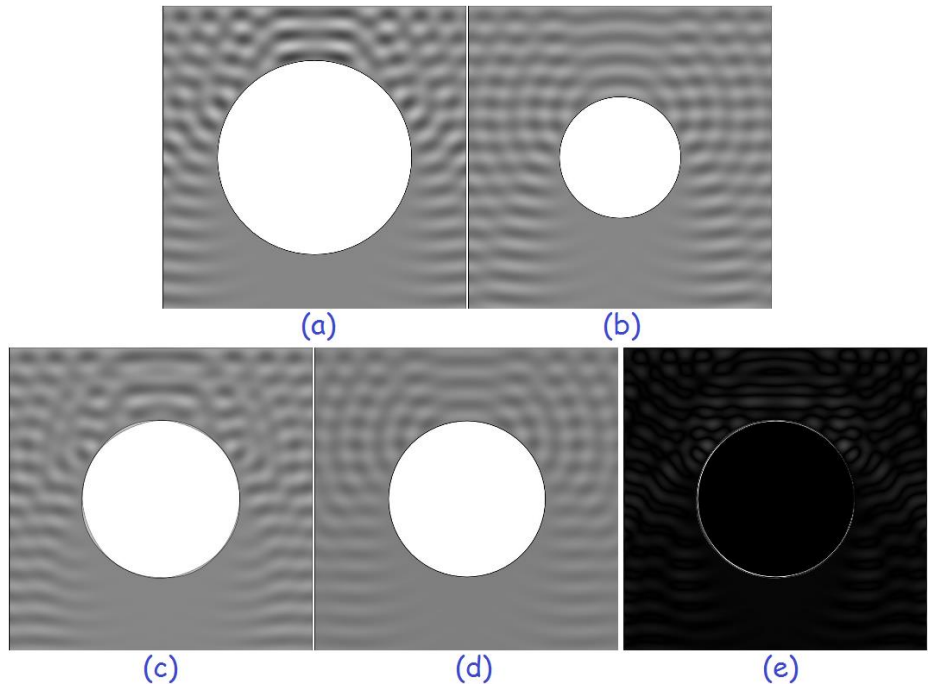

Fig. 8. Infinite conducting obstacle of radii 0.64m (a) and 0.4m (b). Obtained intermediate result (image half-way from the final image) in the case of morphing in (c), and a direct computation in (d). In (e) we show the difference image between (c) and (d). One notes that the scatterer field in (a)-(d) are identical to that in (a)-(d) in Figure 7. On the contrary, panel (e) in Fig. 7 and 8 are much different.

We applied our $L^2$ norm image comparison's function to these difference images in Fig. 7(e) and Fig. 8(e), and we obtained a result close to 0.5 for these two difference images, which means that there is more than 25% of differences between the morphing results and the calculated results. This clearly shows that morphing should be used knowingly, and its range of applicability to transformation optics does not include metamaterials designed via space folding.

*3.5 A morphing step towards a rotacon*

We can also envisage a different use of morphing, for instance, we know concentrators [11], we know rotators [12], but we could imagine a mix of both applications, and above all, how can one model such a combination? A natural way to do it is via morphing.

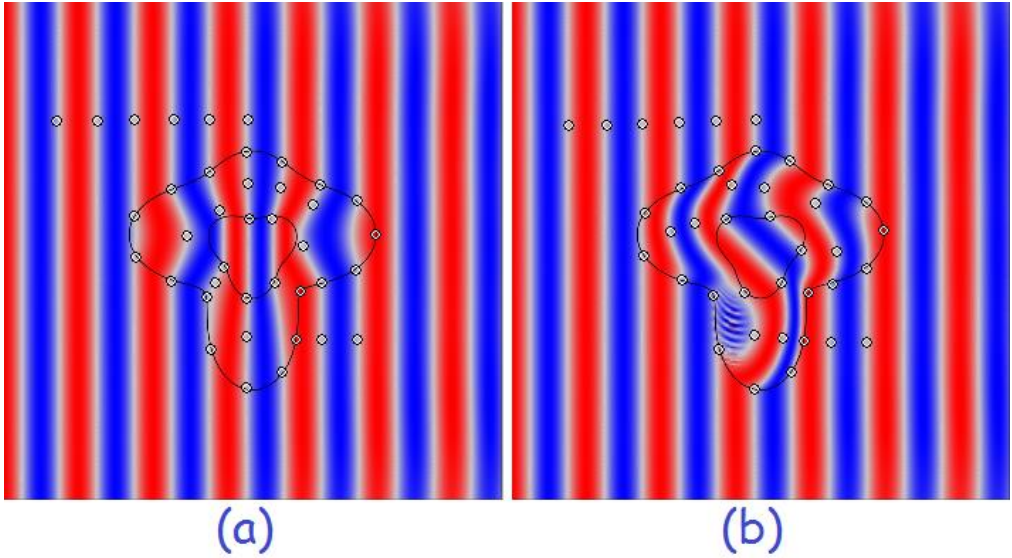

Fig. 9. Non Circular cloak's shape concentrating (a) and rotating (b) the electric field, with their control points.

In order to reproduce the morphing result of the combination of these two functions, namely concentrator and rotator, we introduce the following map:

$$(r,\theta) \xrightarrow{f_I} (r',\theta')$$

$$f_i = \begin{cases} \alpha_i r + \beta_i = r' \\ \theta + \gamma_i r + K_i = \theta' \end{cases} i = II, III$$

$$\forall 0 < r < R_2(\theta) \exists f_{II} / \alpha_{II} = \frac{R_1(\theta)}{R_2(\theta)}, \beta_{II} = 0, \gamma_{II} = 0, K_{II} = 0$$

$$\forall R_2(\theta) < r < R_3(\theta) \exists f_{III} / \alpha_{III} = \frac{R_3(\theta) - R_1(\theta)}{R_3(\theta) - R_2(\theta)}, \beta_{III} = \frac{R_3(\theta)(R_1(\theta) - R_2(\theta))}{R_3(\theta) - R_2(\theta)},$$

$$\gamma_{III} = \frac{\theta_0}{R_2(\theta) - R_3(\theta)}, K_{III} = \frac{R_3(\theta)\theta_0}{R_3(\theta) - R_2(\theta)}$$

$$f_i / i = I : \alpha_I = 1, \beta_I = 0, \gamma_I = 0, K_I = 0 \quad (1)$$

For a complex shape like the one shown in Fig. 9 figure, one can easily understand that it is a delicate matter to calculate the transformed permittivity and permeability of such a combined functionality, and we have done so numerically in COMSOL MUTIPHYSICS. In this figure, the inner and outer boundaries of the concentrator (see panel a) and rotator (see panel b) can be described in a polar coordinate system (r,θ) whose origin is located at the center of the image, in the following manner:

$$R_1(\theta) = 0.4R(1+0.2\sin(3\theta));\ R_2(\theta) = 0.6R(1+0.2\sin(3\theta));\ R_3(\theta) = R(1+0.2(\sin(3\theta)+\cos(4\theta));$$

(2)

with R=0.4. The transformed permittivity and permeability associated with the rotator and concentrator of a complex shape are deduced from the techniques detailed in [13]. The numerical model has been implemented here again in COMSOL. However, the purpose of the present paper is to explore new functionalities combining a local concentration and rotation of the electromagnetic field without perturbing the overall field, using the morphing approach.

As explained in § 2.2, in order to place control points in such quite different images is not easy. We have chosen the "grey" lines (at the frontier between the extrema of wave amplitude) to place these control points. In a first step, we have placed five control points at the inner boundary of the cloak, see fig. 9(a) and (b). It is the most delicate part of the morphing approach since the wave's behavior is very different between the source and destination images in the neighborhood of the cloak's inner boundary. As one can see, the coordinates of these five control points are markedly different between the two images. This can be attributed to the fact that when the inner part of the cloak keeps the same shape (from the source to the destination image), the wave's shape, in this image's part, is totally different. In a second step, we have placed eighteen control points at the cloak's outer boundary. In fact, we have place sixteen control points at the intersection between the "grey" lines (wave wavefronts) and the cloak's outer boundary, and then, we have added two control points shape to harmonize the cloak's transformation with that of the wave. In a third step, we've placed eight control points inside the cloak, at the maxima or the minima of the "grey" lines, to have a coherent transformation in this image's part. Finally, in a last step, we've placed eight control points outside the cloak to ensure that the outside cloak's transformation will not be driven by the inside's one.

The morphing result is immediate, see Fig. 10(c) for what we get half way between the source and destination images in Fig. 10(a) and Fig. 10(b). Note that we removed the boundaries of concentrator and rotator since these are not useful for the control points. However, the fully numerical result obtained from COMSOL differs substantially from what we observe with morphing. This tells us that some work remains to be done towards the modelling of a rotacon, which can be considered as a novel paradigm of transformation optics.

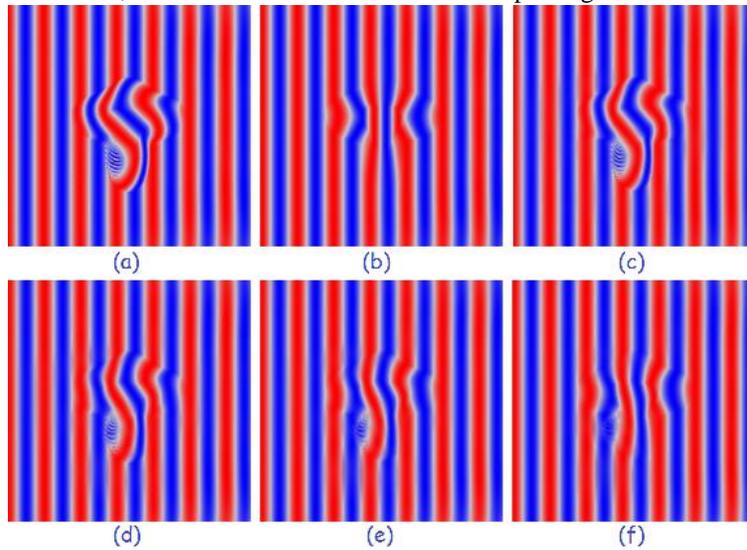

Fig. 10. Morphing result in (c) at 20%, in (d) at 40%, in (e) at 60%, and in (f) at 80%, of a combination of a rotator in (a) with a concentrator in (b) keeping the special cloak's shape of figure 9 (boundaries are not shown as they are not useful in the morphing algorithm here)

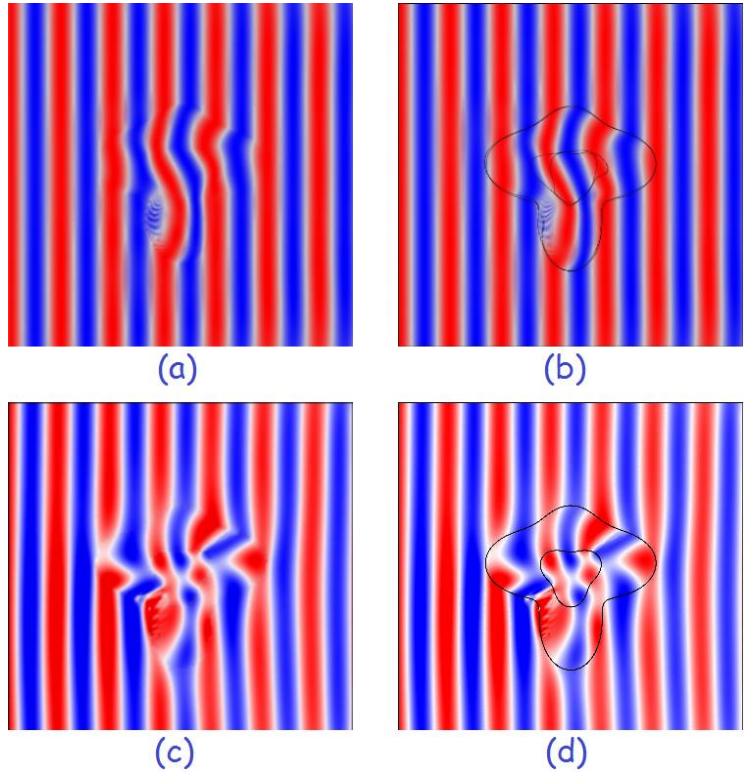

Fig. 11. Morphing result in (a) without boundary, in (b) with boundary of a combination of a rotator with a concentrator keeping the special cloak's shape of figure 9, and numerical result in (c) without boundary, in (d) with boundary of a combination of a rotator with a concentrator keeping the special cloak's shape of figure 9.

**4. Concluding remarks**

A good summary of our work lies in the comparison of the images (b) and (c) in figure 2: The exact and approximated images are virtually indistinguishable by the naked eye.

In this research article, we have explained how morphing works and how to use it, we have seen that it can also unveil new functionalities by mixing two known transformation based metamaterials. One can envisage to apply such an approach for other types of waves, such as acoustic or elastodynamic ones, or to diffusion processes. Morphing may thus prove to be an invaluable tool for the exploration of transformation based metamaterials.

Finally, we would like to point out that while our morphing approach of TO is markedly different from the ray tracing approach of TO discussed in [14], the latter proposal was a source of inspiration for our work.

**Acknowledgements:**

R. Aznavourian and S. Guenneau are thankful for a European funding through ERC Starting Grant ANAMORPHISM.

The authors wish to thank Philippe Réfrégier for his insightful comments on image processing.